\documentclass[pra,letterpaper,twocolumn,showpacs,superscriptaddress,floatfix]{revtex4}

\usepackage{graphicx,psfrag,amsmath,amssymb,amsfonts,bbm,latexsym,color,dcolumn,bm}

\begin{document}

\title{Higgs shifts from electron-positron annihilations near neutron stars}

\author{Gary A. Wegner}

\affiliation{\mbox{Department of Physics and Astronomy, Dartmouth College, 6127 Wilder Laboratory, Hanover, 
NH 03755, USA}}

\author{Roberto Onofrio}

\affiliation{\mbox{Dipartimento di Fisica e Astronomia ``Galileo Galilei'', Universit\`a  di Padova,
Via Marzolo 8, Padova 35131, Italy}}

\affiliation{ITAMP, Harvard-Smithsonian Center for Astrophysics, Cambridge, MA 02138, USA}

\begin{abstract}
We discuss the potential for using neutron stars to determine bounds on the Higgs-Kretschmann 
coupling by looking at peculiar shifts in gamma-ray spectroscopic features. 
In particular, we reanalyze multiple lines observed in GRB781119 detected by two gamma-ray 
spectrometers, and derive an upper bound on the Higgs-Kretschmann coupling that is much 
more constraining than the one recently obtained from white dwarfs. 
This calls for targeted analyses of spectra of gamma-ray bursts from more recent 
observatories, dedicated searches for differential shifts on electron-positron and 
proton-antiproton annihilation spectra in proximity of compact sources, and signals   
of electron and proton cyclotron lines from the same neutron star.
\end{abstract}

\maketitle

\section{Introduction}

The interplay between the mass generation through the Higgs field and gravitation is an 
active subfield of investigation and the concept of mass is crucial both 
in quantum field theory and gravitation. Consequently, any common insight 
may shed light on the possible unification of the two theories.
Inflation, although successful in removing potential contradictions to the standard 
Big Bang model, does not yet have a clear microscopic interpretation and unambiguous 
observational evidence, and the Higgs field has been conjectured to be responsible for it 
\cite{Bezrukov1,Bezrukov2}.
To these long standing motivations, others have been added since the discovery at the LHC 
of a scalar particle with the mass and decay branching ratios as expected from the Higgs 
boson in the minimal Higgs doublet model \cite{Aad,Chatrachyan}. 
In the absence of signals for new physics in the current experimental 
setting, the extrapolation of the standard model to the Planck scale raises 
an issue of stability of the vacuum for the specific value of the quartic 
coupling of the Higgs self-interaction which is extracted from its observed 
mass \cite{Degrassi,Branchina,Buttazzo,Holthausen}.
Various recent contributions point out a solution with no new physics 
beyond the standard model apart from a non-minimal coupling of the Higgs 
field to curvature invariants \cite{Wang,Demir,Torabian}. 
Finally, the tentative interpretation of the BICEP-2 results \cite{BICEP2} in terms of gravitational 
waves produced during inflation, still to be fully scrutinized and compared with 
Planck results \cite{PLANCK}, would call for non-minimal coupling of the Higgs field to gravity 
\cite{Espinosa} or to the inflaton \cite{Fairbairn}.

Recently, we discussed upper bounds to the coupling between the Higgs 
field and a specific curvature invariant, the Kretschmann invariant, based 
on the analysis of molecular lines of $C_2$ and atomic lines of H, C, Ca, 
and Mg from the surface of two white dwarfs \cite{OW}. 
This bound, although already competitive with respect to what is achievable 
in table-top experiments, in principle may be improved by many orders of 
magnitude by exploiting the strong gravity at the surfaces of neutron stars. 
The boost in sensitivity is easily estimated as the Kretschmann invariant 
$K=R_{\mu\nu\rho\sigma}R^{\mu\nu\rho\sigma}$, with $R^{\mu\nu\rho\sigma}$ the Riemann 
curvature tensor, depends on the sixth power of the radius of the astrophysical object.
Thus, for an Earth-radius white dwarf and a 10 km neutron star of equal mass, a gain 
of order $(6000/10)^6 \simeq 4.7 \times 10^{16}$ is expected if spectral features 
could be measured with precision comparable to the ones studied in white dwarfs.
The crucial issue is to get measurable features from neutron star spectra, and in the 
following we discuss possibilities to be considered, in the form of reanalysis of already 
collected data, and future dedicated observations. In Sect. 2 we consider annihilation lines 
tentatively observed during gamma-ray bursts with redshifts compatible to that expected 
from the surface of neutron stars, and discuss the possibility for an anomalous contribution 
with respect to the redshifts due to lines of nuclear origin.
In Sect. 3 this evidence is conservatively considered as a bound on a possible 
signal due to the Higgs shift, and the related bounds discussed. 
In Sect. 4 we discuss possible Higgs shifts arising from a comparative analysis of 
electron-positron and proton-antiproton annihilations, always in proximity of 
a neutron star, which should result in similar bounds to the one discussed earlier. 
More general considerations on the possibility of observing simultaneously cyclotron 
lines from electrons and protons are discussed in the conclusions, as well as possible 
generalizations of the Higgs coupling in theories and models beyond general relativity.

\begin{table}[b]
\begin{center}
\begin{tabular}{l|c|c|c}
\hline\hline
Line & $E_{\mathrm{lab}}(keV)$ & $E_{\mathrm{obs}}(keV)$ & $z_{\mathrm{line}}$ \\
\hline
$e^+e^-$             &  511 &  420 $\pm$ 20  & 0.217 $\pm$ 0.010 \\
${}^{56}$Fe           &  847 &  738 $\pm$ 40  & 0.148 $\pm$ 0.008 \\
${}^{56}$Fe           & 1238 & 1076 $\pm$ 33  & 0.151 $\pm$ 0.005 \\
${}^{24}$Mg           & 1369 & 1164 $\pm$ 36  & 0.176 $\pm$ 0.005 \\
${}^{20}$Ne           & 1634 & 1444 $\pm$ 33  & 0.132 $\pm$ 0.003 \\
${}^{28}$Si           & 1779 & 1589 $\pm$ 33  & 0.120 $\pm$ 0.002 \\
${}^{56}$Fe           & 1811 & 1612 $\pm$ 40  & 0.123 $\pm$ 0.003 \\
${}^{14}$N            & 2313 & 2011 $\pm$ 70  & 0.150 $\pm$ 0.005 \\ 
\hline
\end{tabular}
\end{center}
\caption{\label{table1}
Analysis of the line shifts from the ISEE-3 data for the gamma-ray burst event recorded on 
19 November 1978  reported in \cite{Teegarden2}. The origin of each line is in the first 
column, followed by the transition energy as measured in the laboratory \cite{Kozlovsky}, 
the one observed by ISEE-3, and the related redshift evaluated as $z_{\mathrm{line}}=
E_{\mathrm{lab}}/E_{\mathrm{obs}}-1$. The error bars are evaluated from our analysis of full-width 
half maxima of the interpolating curves appearing in Fig. 6 of \cite{Teegarden2}, and are 
used to obtain weighted average and standard deviation on the eight redshift determinations
for the nuclear lines, resulting in $\langle z_{\mathrm{nucl}} \rangle=0.134 \pm 0.017$, which 
is smaller by $4.2$ standard deviations with respect to the one evaluated using the 
electron-positron annihilation peak, $z_{\mathrm{annih}}= 0.217 \pm 0.010$. 
The instrumental error for the nuclear transitions reported in \cite{Teegarden2} is quoted 
only for the ${}^{56} {\mathrm{Fe}}$ line as 10 $\mathrm{keV}$, corresponding to a relative 
error of $\simeq 1.3 \%$. If we assume the same instrumental error for all remaining nuclear 
lines, then this last is smaller by one order of magnitude with respect to the statistical 
relative error on $\langle z_{\mathrm{nucl}} \rangle$, and we get $\langle z_{\mathrm{nucl}} 
\rangle=0.143 \pm 0.018$, which differs by 3.6 standard deviation from $z_{\mathrm{annih}}$, 
corroborating the former analysis based on our graphical assessment of the error bars in 
Fig. 6 of \cite{Teegarden2}.}
\end{table}

\section{Search for anomalous shifts in gamma-ray bursts}
The most likely avenue towards obtaining better bounds to Higgs-Kretschmann couplings 
from signals in proximity of neutron stars is, to our knowledge, the comparative analysis 
of electron-positron annihilation lines and narrow lines due to nuclear de-excitations 
during transients of gamma-ray bursts. 
In the following we focus on the event collected by the Goddard germanium Gamma-Ray 
Burst Spectrometer on board of ISEE-3, GRB 781119, \cite{Teegarden1,Teegarden2}, as several lines 
attributed to nuclei like ${}^{56}$Fe, ${}^{24}$Mg, ${}^{20}$Ne, ${}^{28}$Si, ${}^{14}$N were identified. 
A less prominent peak at $420$ keV has also been identified and interpreted as 
a red-shifted $e^+e^-$ annihilation peak. The same event was observed by the Konus gamma-ray 
observatory of the Leningrad group  at the Ioffe Institute \cite{Mazets,Golenetskii}, 
including the emission feature at $420$ keV. The redshift required 
to justify this line as a $e^+e^-$ annihilation peak is compatible with the gravitational 
redshift ($z \simeq 0.2$) expected on the surface of a neutron star. 
More qualitatively, the putative electron-positron line is observed at 
$E_{\mathrm{annih}}=(420 \pm 20)$ keV, which is $0.82 \pm 0.04$ times the value of the electron-positron 
two body annihilation line at 511 keV. For the ${}^{56}$Fe line, we have an observed value of 
$E_{\mathrm{Fe}}=(738 \pm 10)$ keV versus an unshifted energy of $847$ keV, {\it i.e.} an energy 
ratio of $0.87 \pm 0.01$. Therefore the electron-positron line is more redshifted than the 
${}^{56}$Fe line by one standard deviation. 
Although the discrepancy between the two redshifts is contained within one standard deviation only, 
by inferring the redshifts of the other identified nuclear lines from Fig. 6 in \cite{Teegarden2} we 
notice that they all result in systematically smaller redshifts than the electron-positron 
annihilation line, at the level of 4.2 standard deviations, as discussed in Table 1. 

\begin{table}[b]
\begin{center}
\begin{tabular}{l|c|c|c}
\hline\hline
Event      & $E_{\mathrm{obs}}(keV)$ & $z_{\mathrm{line}}$ & M/R \\
\hline
18/09/78  & 380 & 0.345 & 0.447 \\
21/09/78  & 350 & 0.460 & 0.531 \\
06/10/78a & 420 & 0.217 & 0.324 \\
06/10/78b & 350 & 0.460 & 0.531 \\      
23/10/78  & 280 & 0.825 & 0.700 \\
05/03/79  & 380 & 0.345 & 0.447 \\
06/04/79  & 320 & 0.597 & 0.608 \\
02/05/79  & 470 & 0.087 & 0.154 \\
26/05/79  & 320 & 0.597 & 0.608 \\
22/06/79  & 450 & 0.136 & 0.224 \\
28/06/79  & 410 & 0.246 & 0.356 \\
09/11/79  & 320 & 0.597 & 0.608 \\
\hline
\end{tabular}
\end{center}
\caption{\label{table2}
Analysis of the line shifts from emission features in the 
70-470 keV interval from the Konus catalogs \cite{Mazetscatalog1,Mazetscatalog2,Mazetscatalog3}. 
The date of the event for each line is in the first column, 
followed by the observed emission energy, the corresponding 
redshift if attributed to the $e^+e^-$ annihilation peak, and 
the $M/R$ parameter expressed in units of $c^2/(2G_N)$.}
\end{table}

While it is tempting to interpret this extra-shift as due to the Higgs shift,  
more conservatively it can be used to determine an upper bound on the associated Kretschmann-Higgs 
coupling, for two reasons. 

The interpretation of the 420 keV peak as due to electron-positron annihilation is not 
solid, and needs to be corroborated by more data. 
There is a general consensus that spectral features in the 300-400 keV region can be 
interpreted as due to a gravitationally red-shifted $e^+e^-$ annihilation line, see 
\cite{Johnson} for observations of the galactic center and \cite{Ramaty} for their interpretation.
However, the absence of similar signals in later observational campaigns like BATSE 
strongly constrains the initial interpretation of the observed peaks \cite{BATSE1,Hurley,BATSE2}. 
On the instrumental side, the energy deposition of each photon may not be completely occurring 
within the detector, and therefore assumptions must be made on the incident spectrum 
\cite{Fenimore,TeegardenWoosley,Loredo}. The very existence of the line features has been 
criticized \cite{Zdziarski,Lamb}, and alternative explanations have been put forward, 
more specifically as origining from de-excitation of ${}^7{\mathrm{Li}}^*$ in cosmic rays 
\cite{Fishman}, and from an amplification mechanism through stimulated annihilation 
radiation \cite{Ramaty1982}. 

Moreover, various environmental factors might create differential shifts 
between the $e^+e^-$ annihilations and the nuclear lines as there are uncertainties 
in the models of gamma ray bursts in neutron stars. 
Although it is reasonable to assume that the nuclear lines originate from matter 
on the surface of the neutron star with null Doppler shift, peculiar motions of 
the $e^+e^-$ plasma clouds may add or subtract a Doppler shift to the gravitational redshift. 
The broadening and shift due to the finite temperature of the $e^+e^-$ plasma cloud could 
reduce the observed redshift. The annihilation line broadens proportionally to $T^{1/2}$ for 
$k_B T << m_e c^2$, and to $T$ for $k_B T >> m_e c^2$, and the peak of the line shifts toward 
higher energies as $\delta E_{\mathrm{peak}}/E_{\mathrm{peak}} \simeq 1.25 k_B T/m_e c^2$ 
\cite{Aharonyan1,Zdziarskishift,RamatyMes,Svensson,Aharonyan2}. 
This blueshift is going to aggravate the redshift excess we have discussed.
It is also worth remarking that a secondary peak on the right side of the putative $e^+e^-$ 
annihilation peak is present at an energy of 484 keV in the ISSE-3 data \cite{Teegarden2}. 
If this is interpreted as the actual $e^+e^-$ annihilation peak, the 
corresponding redshift is only $z_{\mathrm{annih}}=(0.056 \pm 0.002)$, blue-shifted by the average 
value of the redshift from nuclear lines by about 4.3 standard deviations. 
Based on the redshift of the nuclear lines, a temperature of the $e^+e^-$ 
plasma cloud of $\simeq 20$ keV (corresponding to $\simeq 2.6 \times 10^{8}$ K) should 
be required to fully justify this blue-shift, which seems compatible with the typical surface 
temperature of neutron stars.

\begin{figure}[b] 
\vspace{+0.3cm}
\begin{center}
{\includegraphics[width=0.9\columnwidth]{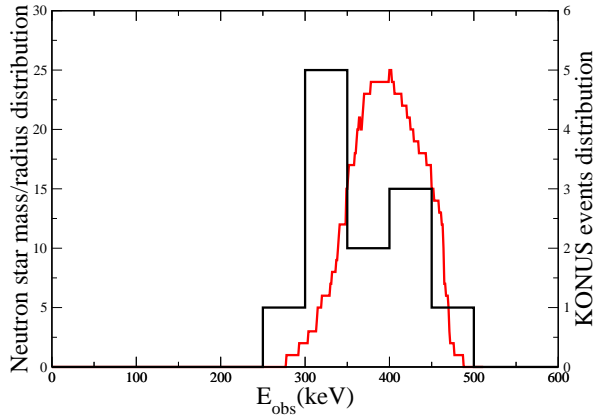}}
\caption{Statistics of the distribution of the observed energy $E_{\mathrm obs}$. 
Distribution of the redshifts ratio as deduced from the emission lines 
of the Konus catalog, see Table 2, in the hypothesis they represent $e^+e^-$ 
annihilation lines (black line, right vertical scale), and from 26 determinations 
of gravitational redshifts for neutron stars \cite{Lattimer1,Lattimer2,Lattimer3} 
(red line, left vertical scale).}  
\label{masstoradiusratio}
\end{center}
\end{figure}

A statistical comparison between the distribution of the energy of the putative $e^+e^-$ 
annihilation peaks in gamma ray bursts and the mass-radius distribution of neutron stars may 
also be used both to check the hypothesis attributing these events to neutron stars and to 
check for systematic deviations. The mass-radius ratio is related to the energy of the 
observed line and the redshift as 
\begin{equation}
\frac{M}{R}=\frac{c^2}{2G_N}\left[1-\left(\frac{E_{\mathrm{obs}}}{E_{\mathrm{lab}}}\right)^2\right]=
\frac{c^2}{2G_N}\left[1-\frac{1}{(1+z)^2}\right].
\end{equation}
The Konus collaboration has published three catalogs of gamma ray bursts 
\cite{Mazetscatalog1,Mazetscatalog2,Mazetscatalog3}, and in 25 cases we find evidence 
for emission peaks at energies in the 100-470 keV range. 
Within this subsample of events, we apply a lower model-dependent cutoff based on the 
requirement for causality \cite{Glendenning,LattimerPrakash}. 
If the events originate on the surface of a neutron star, this 
requires $M/R \leq 0.708 c^2/(2G_N)$ leading to $E_{\mathrm{obs}} \geq 276$ keV.  
The resulting events are reported in Table 2. 
This leads to a statistical distribution of the annihilation energy peaks, depicted in 
Fig. 1, which is also compared with the mass-to-radius ratio distribution 
for a sample of neutron stars obtained from \cite{Lattimer1,Lattimer2,Lattimer3}, 
considering a range of values for $E_{\mathrm{obs}}$ coming from the allowed intervals 
for mass and radius (see \cite{Lattimer1,Lattimer2,Lattimer3} for a discussion 
on the errors in the determination of mass and radius, and their sensitivity to the 
equations of state) and summing up the resulting rectangular windows of unit height. 
The two distributions are at least limited by roughly the same interval, but 
the Konus distribution seems more peaked at lower energies. The small number 
of events in the Konus data prevents us to make more quantitative 
analyses\footnote{We have analyzed the low energy spectra of the Konus-Wind 
data for the year 2003 publicly available at the NASA website: 
http://asd.gsfc.nasa.gov/konus/, with no evidence for spectral features apart from 
a recurrent peak at about 260 keV with approximately constant height, presumably of 
instrumental origin.}.

\section{Bounds on the Higgs-Kretschmann coupling}

As detailed in former contributions \cite{OW,Onofrio,Onofriograv}, if the Higgs field 
$\phi$ characterized by quadratic and quartic coefficients $\mu$ and $\lambda$ is 
coupled to the Kretschmann curvature invariant $K$ via the Lagrangian density term  
$\xi_K  \Lambda_\mathrm{Pl}^2 \phi^2 K$, with $\xi_K$ their coupling constant and 
$\Lambda_\mathrm{Pl}$ the Planck length, the effective mass parameter 
of the Higgs field gets an extra-term due to the scalar curvature as 
$\mu^2 \mapsto \mu^2(1+\xi_K \Lambda_\mathrm{Pl}^2 \lambda_{\mu}^2 K)$, with 
$\lambda_\mu$ the Compton wavelength of the Higgs field corresponding to its mass of 125 GeV
($\lambda_{\mu}=1.6 \times 10^{-18}$ m), and the vacuum expectation value of the Higgs field $v$ 
depends on space as:
\begin{equation}
v=\sqrt{-\frac{\mu^2+\xi_K \Lambda_\mathrm{Pl}^2 K}{\lambda}} \simeq v_0 
\left(1+\frac{\xi_K \Lambda_\mathrm{Pl}^2 K}{2 \mu^2}\right),
\end{equation}
in a weak-curvature limit, {\it i.e.} with curvature length scale much larger than 
$\lambda_\mu$, a limit well satisfied in all circumstances of astrophysical interest. 
Then the mass $m_e$ of the electron will be 
simply changed proportionally to the Higgs vacuum expectation value 
\begin{equation}
\delta m_e=\frac{y_e}{\sqrt{2}} (v-v_0)\simeq 
\frac{y_e \xi_K \Lambda_\mathrm{Pl}^2 K v_0}{2^{3/2}\mu^2} = 
\frac{1}{2} \xi_K \Lambda_\mathrm{Pl}^2 \lambda_\mu^2 K m_e.  
\end{equation}
where $y_e$ is the Yukawa coupling of the electron. 
Hadrons, having most of their mass arising from QCD vacuum, are instead 
minimally affected by the curvature invariant.
In order to get upper bounds based on the data above, we first check that the 
Higgs-shift expected for the nuclear lines is indeed negligible with respect to the one 
expected in the $e^+e^-$ annihilation peak. 
The relationship between the mass of a nucleon and the energy levels of the corresponding 
nucleus is not trivially available, as there are many phenomenological models based either 
on a single nucleon approach, in which a nucleon evolves in the mean-field potential created 
by the remaining nucleons, as in the shell model approach, or on a many-body collective 
approach as in the droplet model \cite{Caurier}. Considering that strong interactions 
are mainly responsible for the binding between nucleons, we assume that in a simplified 
treatment with a harmonic oscillator potential, the energy levels will scale as the 
inverse of the square root of the involved mass, the mass of the nucleon in the shell 
models, and the whole mass of the nucleus in collective models. 
In this case the expected Higgs shift should scale as 
$\delta \lambda/\lambda \simeq  \delta m_n/(2m_n)$, with $m_n$ the relevant mass 
(ranging between the two extreme values of the nucleon mass, for instance the one of 
the proton $m_p$, and the nucleus mass $m_N$). In an infinite square-well model instead 
the expected scaling should be the inverse of the mass, $\delta \lambda/\lambda \simeq 
\delta m_n/m_n$, a mere factor 2 larger than in the harmonic potential.

In the case of a single nucleon, relevant for single-particle models such as the shell model,  
we have, focusing on the proton mass, $m_p \simeq m_{QCD}+(2 y_u+y_d)v/\sqrt{2} \simeq 928+ 10 
(1+0.5 \xi_K \Lambda_P^2 \lambda_{\mu}^2 K)$ (with $y_u$ and $y_d$ the Yukawa couplings of the 
up and down quarks, $m_{QCD}$ the purely gluonic contribution to the proton mass, and 
all masses and energies expressed in MeV/c${}^2$), which implies 
$\delta m_p/m_p \simeq 5 \times 10^{-3} \xi_K \Lambda_P^2 \lambda_\mu^2 K$. 
This has to  to be compared with the sensitivity to the Kreschmann-Higgs 
coupling of the electron, $\delta m_e/m_e \simeq 0.5 \xi_K \Lambda_P^2 \lambda_\mu^2 K \simeq 
10^2 \delta m_p/m_p$.
In order to estimate the mass shift in the case of collective models, we consider the nucleus 
with the larger number of lines observed as in Table 1, the iron isotope with mass number 56 
made of 26 protons and 30 neutrons. Its mass can be written in terms of the Yukawa couplings 
$y_u$ and $y_d$ of the up and down valence quarks inside protons and neutrons obtaining

\begin{equation}
m_{\mathrm{Fe}}=56 m_{\mathrm{QCD}}+\frac{(82 y_u+86 y_d)v}{\sqrt{2}}-\Delta m(A,Z),
\end{equation}
where we have introduced the mass defect $\Delta m(A,Z)$. In the presence of a curved 
spacetime with Kretschmann coupling to the Higgs field the mass of a ${}^{56}$Fe nucleus will be

\begin{equation}
m_{\mathrm{Fe}}(\xi_K)=m_{\mathrm{Fe}}(0) \left(1+\frac{82y_u+86y_d}{2\sqrt{2}}\xi_K \Lambda_P^2 
\lambda_\mu^2 K \right).
\end{equation}
Again, the relative mass shift of the nucleus turns out to be smaller than the relative 
mass shift of the electron by a factor even larger than in the single nucleon case, since 
$(\delta m_e/m_e)/(\delta m_{\mathrm{Fe}}/m_{\mathrm{Fe}}) 
\simeq \sqrt{2}m_{\mathrm{Fe}}/(82 y_u+86 y_d)v_0 
\simeq 87$. 
Therefore, both the extreme examples of the single nucleon mass and the whole ${}^{56}$Fe mass 
determining the nuclear spectroscopy show that their contributions are negligible with respect 
to the mass shift of the electron, the nuclear line thus providing a spectroscopic `anchor'. 
This implies that we may attribute the Higgs shift to the electron mass shift alone and, by assuming 
the Planck length $\Lambda_P=10^{-35}$ m and a solar mass neutron star with radius 10 km (corresponding 
to a Kretschmann invariant on the neutron star surface of $K=10^{-16} \mathrm{m}^{-4}$), the maximum 
Higgs shift compatible with the observed excess of redshift 
$\delta z=z_{\mathrm{annih}}- \langle z_{\mathrm{nucl}} \rangle\geq 1.28 \times 10^{-122} \xi_K$, 
gives an upper  bound on the Higgs-Kretschmann coupling $\xi_K \leq 5.8 \times 10^{120}$ 
in MKSA units, translated into a value of $7.1 \times 10^{35}$ in natural units. 
If the line at 484 keV is instead considered responsible for the $e^+e^-$ annihilation, 
the Higgs shift is negative, 
$\delta z=z_{\mathrm{annih}}- \langle z_{\mathrm{nucl}} \rangle=-0.087 \pm 0.018$, 
corresponding to an upper bound of  $\xi_K \geq -6.8 \times 10^{120}$ ({\it i.e.} 
$|\xi_K|\leq 6.8 \times 10^{120}$), always in MKSA units.
By assuming $\Lambda_P=10^{-19}$ m as in models with the Planck scale coinciding with 
the Fermi scale \cite{Arkani1,Arkani2}, the bounds are correspondingly stronger by 
a factor $\simeq 10^{32}$.

\begin{figure}[b] 
\vspace{+0.3cm}
\begin{center}
{\includegraphics[width=0.9\columnwidth]{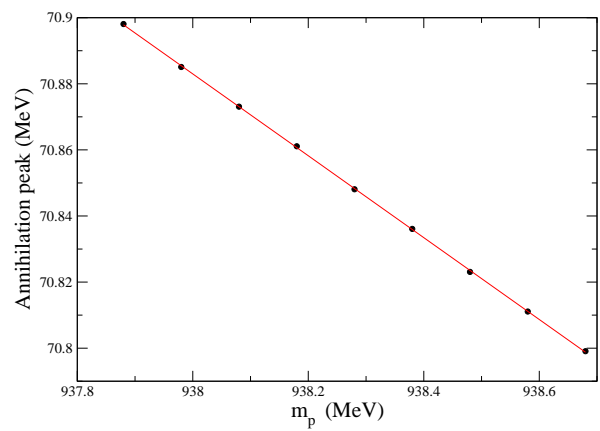}}
\caption{Sensitivity of the energy distribution peak in $p\bar{p}$ annihilations 
to the proton mass. The fit yields $\bar{E}_\gamma=187.05 - 0.12385 m_p$, with 
energies and the proton mass expressed in MeV.}  
\label{HiggsNSfig2}
\end{center}
\end{figure}

\section{Electron-positron and proton-antiproton annihilations near neutron stars}

An alternative possibility to study Higgs shifts is to compare electron-positron and 
proton-antiproton annihilation signals from neutron stars, as suggested in \cite{Onofrio}. 
Unlike $e^+e^-$ annihilations, $p\bar{p}$ annihilations produce a continuous photon 
spectrum since the annihilation produces multiple $\pi^0$ mesons in turn decaying into photons. 
Consequently the photon spectrum has an intrinsically broad peak due to the 
more than two-body decays, which is further Doppler-broadened by the velocity spread of 
the involved particles. Due to the flatness of the expected photon spectrum from $p\bar{p}$ 
annihilation, possible limits on the Kretschmann coupling from this class of events are relatively 
weaker than the former class of events. The continuous gamma-ray spectrum in the $p\bar{p}$ 
annihilations has been evaluated in \cite{Backenstoss} by fitting a Monte Carlo simulation 
with an analytical function as

\begin{equation}
F(E_\gamma)=N [(E_p-E_\gamma)^{\alpha_1}e^{\beta_1}+(E_p-E_\gamma)^{\alpha_2} e^{\beta_2}+\beta_3 e^{\alpha_3 E_\gamma}],
\end{equation}
where $E_p=m_p c^2$ is the proton mass in energy units, and $\alpha_i, \beta_i$ ($i=1 \div 3$) are 
fitting parameters available in \cite{Backenstoss}. 
The position of the annihilation peak depends upon the assumed proton mass, as shown in Fig.2, 
with a best fit yielding $\bar{E}_\gamma=187.05 - 0.12385 E_p$ (all energies in MeV), with 
the slope $\delta \bar{E}_{\gamma}/\delta m_p =0.12385$ expressing its sensitivity. 
If the minimum detectable peak shift is then $\delta \bar{E}_\gamma \simeq 0.1$ MeV, 
the minimum detectable proton mass shift is $\simeq 0.8$ MeV, {\it i.e.} 
$\delta m_p/m_p \simeq 8 \times 10^{-4}$, and the situation 
is similar to the one in the previous section as the sensitivity of the proton shift 
to the spacetime curvature is far smaller than the one of the electron. 
For the electron-positron annihilations, the limitation is due to the intrinsic 
resolution of the 511 keV peak which depends upon the environmental temperature 
and energy resolution of the detector, estimated, respectively, to be  
$\Delta E/E\vert_{\mathrm{env}} = K_B T/E_\gamma \simeq 10^{-5}$, and  
$\Delta E/E\vert_{\mathrm{instr}}=1.47 \times 10^{-4}$ \cite{Churazov}. 
By using the Rayleigh criterion for resolving a shift of the annihilation peak, with a 
full-width half maximum of 2.37 keV as quoted in \cite{Churazov}, and the same values 
of $\Lambda_{\mathrm{Pl}}$, $\lambda_{\mu}$, mass and radius of the neutron star used above, 
we get a bound $\xi_K=4.4 \times 10^{34}$ using natural units, one order of magnitude 
stronger than to the one evaluated above for the gamma-ray burst event. 
Table 3 summarizes our discussion by including upper bounds from laboratory measurements, 
actual spectroscopy from white dwarfs, and the potential observations from neutron stars 
reported in this paper.

\begin{table}
\begin{center}
\begin{tabular}{l|c}
\hline\hline
Source & $\xi_K (\mathrm{\Lambda_{Pl}=10^{-35} m})$  \\
\hline
Table-top Experiments                        &  $2.5 \times 10^{60}$  \\
BPM 27606                                    &  $5 \times 10^{50}$    \\
Procyon B                                    &  $9 \times 10^{50}$    \\
$e^+e^- \mathrm{vs. nuclei} {\mathrm{(a)}}$   &  $7.1 \times 10^{35}$  \\
$e^+e^- \mathrm{vs. nuclei} {\mathrm{(b)}}$   &  $-8.4 \times 10^{35}$ \\
$e^+e^- \mathrm{vs.} p\bar{p}$                &  $4.4 \times 10^{34}$  \\
\hline
\end{tabular}
\end{center}
\caption{\label{table3}
Summary of bounds on the Higgs-Kretschmann couplings (in Natural Units) from various sources, 
experiments to test the superposition principle of the gravitational force as discussed 
in \cite{Onofriograv}, the analysis of differential shifts in spectral lines from two white 
dwarfs \cite{OW}, the analysis of differential shifts between $e^+e^-$ annihilation 
line and nuclear lines with two candidates at 420 keV (a) and 484 keV (b) as discussed in Sect. 3, 
and possible comparisons between $e^+e^-$ and $p\bar{p}$ annihilation lines originating from the 
surface of neutron stars, as discussed in Sect. 4. The bounds are evaluated for a choice of the 
Planck length of $\Lambda_{\mathrm{Pl}}=10^{-35}$ m, the bounds for the choice of 
$\Lambda_{\mathrm{Pl}}=10^{-19}$ m being $10^{32}$ times stronger.}
\end{table}

\section{Conclusions}

In summary, we have discussed neutron stars as potential tools to constrain a specific 
Higgs-curvature connection. The most promising seems to be a reanalysis of the redshifted 
signals during GRB events, and analysis of recent data taken with gamma-ray observatories 
could be targeted looking for this peculiar effect.
This suggests the need for a comprehensive reanalysis of gamma-ray bursts in which transient 
features appear in the energy spectra. There is tension between the various observational 
parameters involved, as one simultaneously makes three demands: high time resolution 
to avoid washing out the transient in case of a sampling time too large, high energy resolution 
to identify with enough precision the location of the lines, and large statistics to avoid 
the signal being immersed in the background. This also adds motivations to the development 
of satellite detectors in the 1 keV-10 MeV range with high spectral and temporal resolutions, and 
large gamma spectrometers on balloons \cite{Fishman1,Fishman2,Fishman3,Meegan} in which the 
shorter observation time could be offset by the larger fiducial detection volume, or the 100-day 
observation time planned for the Ultra Long Duration Balloon program 
\cite{Cathey,Fuke,Aramaki,Website}. An alternative method could be the simultaneous 
observation of $e^+e^-$ and $p\bar{p}$ annihilations, and we have shown that bounds 
are of similar order of magnitude if state of art instrumental resolution can be achieved. 

We also mention the feasibility of observations of cyclotron lines of electrons 
and protons or ions in the same region of magnetic fields of neutron stars. 
So far there have been observations of both lines but in different neutron stars, 
see \cite{Ibrahim} for the evidence of a line feature in a soft gamma repeater 
interpreted as a proton cyclotron resonance, \cite{Tiengo} for a feature from a 
magnetar also interpreted as a proton cyclotron feature, and  \cite{Bignami} for 
a band of electron cyclotron lines from an isolated neutron star. 
The mismatch of the two cyclotron frequencies by the proton-to-electron mass 
ratio makes their simultaneous observation on the same neutron star quite difficult. 
This mismatch is smaller for electron and proton spin-flip resonances, however, qualitative 
estimates show that their absorption signal is suppressed, with respect to the one due to 
the cyclotron resonance, by a factor approximately equal to the fine structure constant, ruling 
out its observability with the current data \cite{Melrose,Thompson,Zane}.

The work discussed so far on Higgs shifts based on a Kretschmann coupling can also be 
extended in the analysis of bounds to Higgs-curvature couplings with the Ricci scalar
for models beyond general relativity\footnote{We need to point out that, according to the 
discussion presented in \cite{Okon}, the presence of nonminimal couplings between spacetime 
curvature and any quantum field gives rise to violations of the equivalence principle
putting the models outside the realm of general relativity.}. 
The Jebsen-Birkhoff theorem, {\it i.e.} the fact that the Schwarzschild solution is the 
unique spherically symmetric vacuum solution, does not hold in metric $f(R)$ gravity, 
and $R \neq 0$ even if $T=0$ (see \cite{Goenner} for a dedicated study of possible counterexamples). 
This means that a putative observed Higgs shift will be differently interpreted in various 
approaches, as due to the Kreschmann-Higgs coupling in ordinary general relativity, of 
as a Ricci-Higgs coupling in higher dimensional theories \cite{Wehus1,Wehus2}, $f(R)$ 
theories \cite{Faraoni,Nzioki}, or in Ho\v{r}ava gravity \cite{Horava1,Horava2,Mazzitelli}. 
Further observables will therefore be necessary to disentangle the various theoretical scenarios. 

%\acknowledgments

\begin{acknowledgements}
We are grateful to J.M. Lattimer for useful correspondence, and to 
S. Lenzi , R. M. Millan, and R. Turolla for helpful discussions.
\end{acknowledgements}

\end{document}